\documentclass[a4paper,11pt]{revtex4}
\usepackage{amsmath}
\usepackage{amssymb}
\usepackage{graphicx}
\usepackage[hypertex]{hyperref} % for hyperlinks between label-ref bibitem-cite

\usepackage[utf8]{inputenc} % entree 8 bits 
\usepackage[T1]{fontenc}      % encodage 8 bits des fontes utilisees
\usepackage{textcomp}            % symboles complémentaires (euro)
   \usepackage[frenchb]{babel}      % titres en français

\def\eps{{\varepsilon}}

\def\atanh{{\rm {atanh}}}
\newcommand{\al}{\alpha}
\newcommand{\hex}{h_{\mathrm{ext}}}

\begin{document}

\title{The L\'evy spin glass transition}
\author{K. Janzen$^{1}$, A. Engel$^{1}$ and M. M\'{e}zard${}^{1,2}$}
\affiliation{
${}^1$ Institut f\"ur Physik, Carl-von-Ossietsky Universit\"at, 26111
  Oldenburg, Germany 
\\
${}^2$Laboratoire de Physique Th\'eorique et Mod\`eles Statistiques,
CNRS and Universit\'e Paris-Sud, B\^{a}t 100, 91405 Orsay, France
}

\begin{abstract}
We determine  the phase transition of the L\'evy spin glass. A
regularized model where the coupling constants smaller than some
cutoff $\eps$ are neglected can be studied by the cavity method for
diluted spin glasses. We show how to handle the $\eps\to 0$ limit and
determine the de Almeida-Thouless transition temperature in presence
of an external field. Contrary to previous findings, in zero external
field we do not find any stable replica-symmetric spin glass phase:
the spin glass phase is always a replica-symmetry-broken phase.  
\end{abstract}

\maketitle

The L\'evy spin glass, introduced in \cite{CB}, is a mean field spin
glass model where  the distribution of couplings has a power law tail
with a diverging second moment. This model can be useful to study some
experimental situations (like metallic spin glasses with RKKY
interactions), it also provides a situation which is intermediate
between the SK model \cite{SK} and finite connectivity mean 
field spin glasses \cite{VB,MePa86,KaSo,MezParBethe}. It is
 particularly relevant for the study of the importance of rare, but
 strong, coupling constants. In their pioneering work, Cizeau and
 Bouchaud \cite{CB} have computed the  spin glass transition
 temperature, and argued that these rare and strong couplings can
 stabilize a  replica symmetric (RS) stable spin glass phase in the
 absence of external magnetic field. Here we  revisit this problem
 using the RS cavity method. We show that replica symmetry is always
 broken in the spin glass phase, and we compute the de Almeida
 Thouless (AT) line \cite{AT} giving the phase diagram as function of
 temperature and   magnetic field.

We consider an Ising spin glass with Hamiltonian  
\begin{equation}
  \label{eq:H}
  H(\{S_i\})=-\frac{1} {2} \sum_{(i,j)} J_{ij} S_i S_j-\hex \sum_i S_i\; ,
\end{equation}
where the sum is over all pairs of spins $S_i=\pm 1,\, i=1,...,N$ and
$\hex$ denotes an external field. The couplings $J_{ij}=J_{ji}$ are
independent, identically distributed random variables drawn from the
distribution  
\begin{equation}
  \label{eq:defP}
 P_\al(J)=\frac{\al}{2 N}\frac{1}{|J|^{\al+1}}\;\theta(|J|-N^{-1/\al})\; , 
\end{equation}
where $\theta(x)$ denotes the Heaviside function and $\al\in]1,2[$ is a
parameter. The coupling distribution is dominated by its power law
tails which are equivalent to those of a L\'evy distribution with parameter
$\al$ \cite{GnKo}. The scaling of the couplings with $N$ ensures that
the free energy corresponding to the Hamiltonian (\ref{eq:H}) is
extensive \cite{CB,JHE}.  

The equilibrium thermodynamic properties of the system at temperature
$1/\beta$ can be deduced from the probability distribution $P(h)$ of
local fields $h_i$ parametrizing the marginal distribution of spin
variables by $P(S_i)=e^{\beta h_i S_i}/2\cosh(\beta h_i)$. Adding a
new site $i=0$ with corresponding couplings $J_{0i}$ to the 
system the new field $h_0$ is given by \cite{MezMonBook}
\begin{equation}
h_0=\hex+\sum_{i=1}^N u(h_i,J_{0i}) 
\label{eq:update}
\end{equation}
where $u(h,J)=\atanh\big(\tanh(\beta h) \tanh(\beta J)\big)/\beta$. 

This relation can be turned into a self-consistent equation for $P(h)$
by averaging over $h_i$ and $J_{0i}$. Within the assumption of replica
symmetry the $h_i$ are independent and we find in the thermodynamic 
limit $N\to\infty$  
\begin{align}\nonumber
 P(h)&=\int\!\!\prod_i dh_i P(h_i)\!\!\int\!\! \prod_i dJ_{0i} P_\al(J_{0i})\;
  \delta(h-\hex-\sum_{i=1}^N u(h_i,J_{0i}))\\
     &\to \int\frac{ds}{2\pi} \exp{\Big[is(h-\hex)+\frac{\al}{2}
        \int dh' P(h')  \int \frac{dJ}{|J|^{\al+1}}
          \big(e^{-is u(h',J)}-1\big)\Big]}  
\label{sceq} 
\end{align}
For $\hex=0$ this equation is equivalent to the one obtained in
\cite{JHE} using the replica method. 
Notice that the result is universal: the  field distribution $P(h)$
depends only on the L\'evy tail of the distribution of couplings
$P_\alpha$, not on the precise definition of its cutoff at small $J$.

It is instructive to solve (\ref{sceq}) numerically with a  population
dynamics \cite{MezParBethe}  method. In order to do this, one should
first realize that in the update equation (\ref{eq:update}) the main
contribution is obtained from the relatively rare  couplings which are
finite in the large $N$ limit. Let us introduce a threshold $\eps$ and
divide the couplings into strong ($|J_{ij}|>\eps$) and weak 
($|J_{ij}|\leq\eps$)  couplings. Eq.  (\ref{eq:update}) involves a sum over
${\cal  O}(\eps^{-\al})$ strong couplings, which is treated exactly,
and a sum over ${\cal  O}(N)$  weak ones, which is  approximated by a
Gaussian random variable $z$ with zero mean and a variance determined
self-consistently. The resulting population dynamics algorithm is
given by \cite{JEM}:
\begin{align}\nonumber
 h_j&=\hex+\sum_{k=1}^K  u(h_k,J_k) + z\\
 \overline{z^2}&=\al\int dh P(h) \int_0^\eps
   \frac{dJ}{J^{\al+1}} \;u^2(h,J)
\label{hlargesmall}
\end{align}
where $K$ is a Poissonian with average $\eps^{-\al}$. In this form the
algorithm represents a noisy variant of the one used for locally
tree-like graphs \cite{MezParBethe}.

\begin{figure}
  \begin{minipage}[c]{.5\textwidth}
  \includegraphics[width=\textwidth]{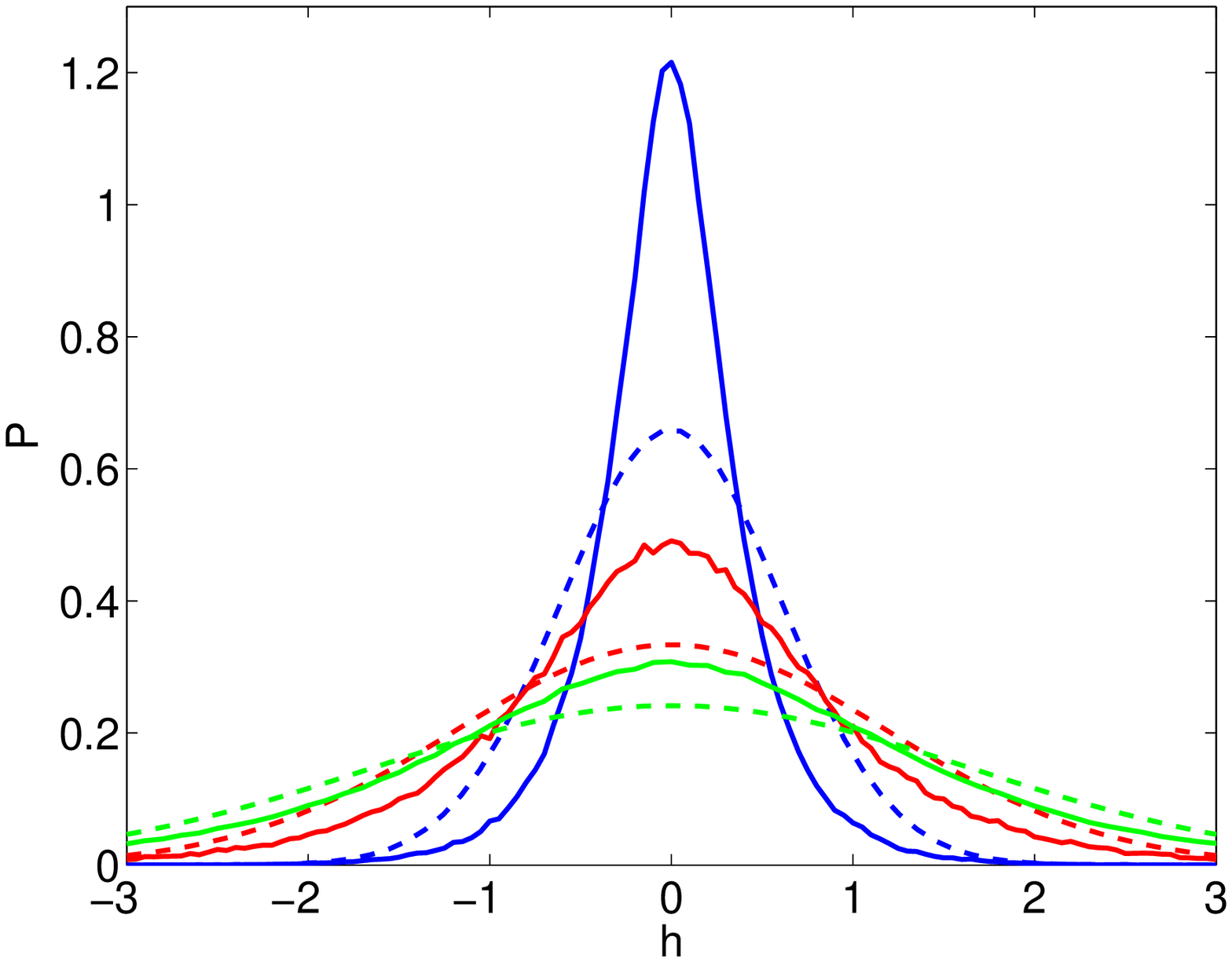}
\end{minipage}%
 \begin{minipage}[c]{.5\textwidth}
 \includegraphics[width=\textwidth]{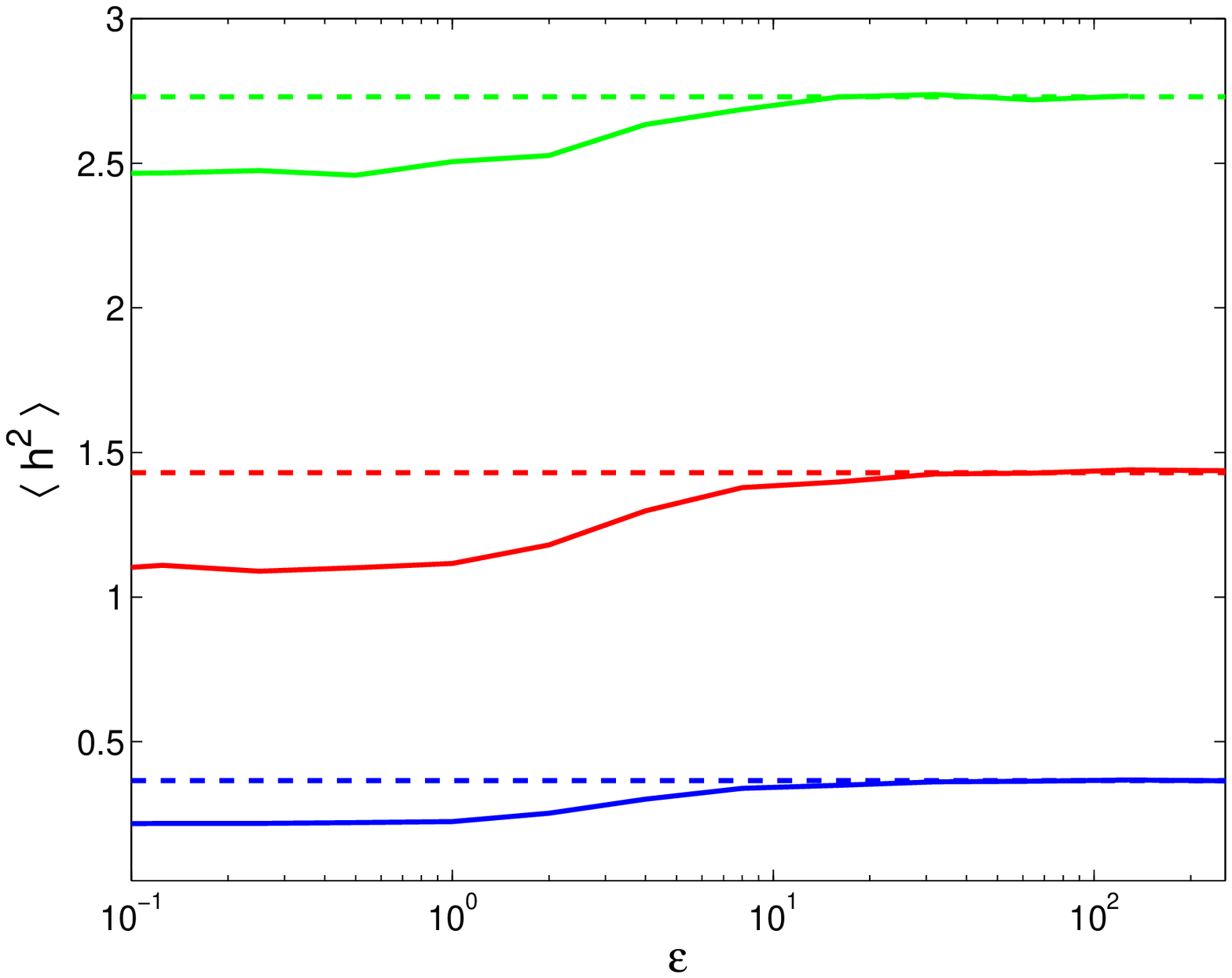}
\end{minipage}
  \caption{Distribution of local fields $P(h)$, for $\alpha=1.1$ and
  $T=0.9\; T_c$ (blue), $T=0.6\; T_c$ (red), and $T=0.1\; T_c$
  (green). The dotted lines are the corresponding results within the
  Gaussian approximation proposed in \cite{CB}. Left: $P(h)$ obtained
  from population dynamics with $\eps=0.3$. Right: The second moment
  of  $P(h)$, obtained with the population dynamics method as function
  of the regularization parameter $\eps$.} 
\label{fig12}
\end{figure}
When $\hex=0$, one easily finds that $P(h)=\delta(h)$ for
$T>T_c(\alpha)$, and $P(h)$ becomes non-trivial at
$T<T_c(\alpha)$. The spin glass transition temperature $T_c(\alpha)$
is independent of $\eps$, it is given by 
\begin{equation}\label{Tc}
T_c(\alpha)=\left[ \int_0^\infty \frac{\al\,dx}{x^{\al+1}} 
   \tanh^2 x\right]^{\frac{1}{\al}} 
\end{equation}
as found in \cite{CB,JHE}.

It is interesting to study the dependence of the resulting 
distributions $P(h)$ on $\eps$. The correct (within the RS
approximation) $P(h)$ is obtained in the limit $\eps\to 0$ whereas the
limit $\eps\to \infty$ amounts to approximating $P(h)$ by a Gaussian,
as in \cite{CB}.  We show in Fig.~\ref{fig12} the second moment
$\langle h^2\rangle$ of $P(h)$ as a 
function of $\eps$.  The convergence of $\langle h^2\rangle$ in the
limit $\eps\to 0$ is very smooth and results obtained for
$\eps=0.2\dots 0.5$ are already very good approximations for the exact
value at $\eps=0$. A truncated model where the weak bonds
are completely neglected would also give the correct result when
$\eps\to 0$, but the convergence is much faster with our
self-consistent Gaussian approximation for the weak bonds.  

In zero external field, we have checked the result  by a direct
iteration of  (\ref{sceq})  using the fast Fourier transform.
Fig.~\ref{fig12} shows some examples of  $P(h)$, together with the
Gaussian form proposed in \cite{CB}. It is clear that $P(h)$ deviates
from a Gaussian distribution. This can already be seen from
(\ref{sceq}): inserting a Gaussian $P(h')$ in the r.h.s. does not
produce one in the l.h.s.  

%%%%%%%%%%%%%%%%%%%%%%%%%%%%%%%%%%%%%%%%%%%%%%%%%%%%%%%%%%%%%%%%%%%%%%%%%

We now turn to the computation of the AT line, characterized by
replica symmetry breaking (RSB). The RS cavity method described above
is valid as long as the spin glass susceptibility,  
$\chi_{SG}=\sum_{i,j} \overline{\left( \langle s_i
  s_j\rangle-\langle s_i\rangle\langle s_j\rangle\right)^2}/N$, 
 is finite. The divergence of $\chi_{SG}$ signals the appearance of
 the spin glass phase. In order to compute this susceptibility, we use
 the truncated model where we keep only the strong bonds with
 $|J_{ij}|>\eps$, while the weak bonds are neglected. The exact value
 of $\chi_{SG}$ is obtained  in the limit $\eps\to 0$. In the
 truncated model, the graph of interacting spins is a diluted
 Erd\"os-Reny\'i random graph:  in the $N\to\infty$ limit (taken before
 the $\eps \to 0$ limit),  the number of spins interacting with a
 given spin is a  Poissonian random variable with mean
 $\eps^{-\alpha}$. This graph is locally tree-like, in the sense that,
 if one looks at all the spins at distance $\leq r$ of a given spin
 $s_i$, their interaction graph is typically, in the large $N$ limit,
 a tree of depth $r$. This allows to compute  $\chi_{SG}$ as
  \cite{MezParBethe,MezMonBook}: 
\begin{equation}
\chi_{SG}=\sum_{r=1}^\infty \eps^{-\alpha r} C_2(r)
\label{eq:chiseries}
\end{equation}
where $C_2(r)$ is the average square correlation,  $\overline{\left(
  \langle s_i s_j\rangle-\langle s_i\rangle\langle
  s_j\rangle\right)^2}$, between two sites $i,j$ at distance $r$. As
we will see, $C_2(r)$ decays exponentially with distance as $C_2(r)= A
e^{-r/\xi}$. We thus define the stability parameter
$\lambda=\eps^{-\alpha} e^{-1/\xi}$. This parameter is the rate of the
geometric series (\ref{eq:chiseries}) giving $\chi_{SG}$. The 
 spin glass
phase transition is given by the condition $\lambda=1$. 

Because of the locally-tree-like structure of the interaction graph in
the truncated model, the computation of $\lambda$ reduces to the study
of a one dimensional L\'evy spin glass model, with energy given by:
\begin{equation}
E=-\sum_{n=1}^{r-1} J_n s_n s_{n+1}-\sum_{n=1}^r h_n s_n\ ,
\label{eq:levy1d}
\end{equation}
where the couplings $J_n$ are independent random variables drawn from
the distribution 
$P_{\al,\eps}(J)=\alpha \eps^\alpha/(2 |J|^{1+\alpha}) \theta(|J|-\eps)$,
and $h_n$ are independent random variables drawn from the distribution
of  cavity fields $P(h)$ determined above. $C_2(r)$ is the {\it
  average} square correlation  
 $\overline{\left( \langle s_1 s_r\rangle-\langle s_1\rangle\langle
  s_r \rangle\right)^2}$, and one is interested in computing the decay
rate $1/\xi=-\lim_{r\to \infty} \log(C_2(r))/r$. While this
one-dimensional system looks  simple, it requires some special care.  
The usual 'population approach' used in finite connectivity spin
glasses (\cite{MezMonBook,MezParBethe,KKJ}) fails in the L\'evy
 case, because the ratio between the average and the typical correlation
diverges in the small $\eps$ limit: the well known 'non-self-averageness'
 of correlation functions  \cite{DerridaHilhorst} becomes 
crucial in this case. This fact is most easily seen in the case where
the fields $h_n$ are equal to zero. As we have seen, this  happens
when $\hex=0$ and $T>T_c(\al)$. The average correlation is
$C_2(r)=\left(\int dJ  P_{\al,\eps}(J) \tanh^2 (\beta J )\right)^r$;
in the limit where  $\eps$ goes to $0$ this gives 
$e^{-1/\xi}=\eps^\alpha \int_0^\infty (\alpha dJ
 /J^{1+\alpha})\; \tanh^2(\beta J)$. Therefore  the stability
 parameter is $\lambda =  \int_0^\infty (\alpha dJ /J^{1+\alpha}) \;
 \tanh^2(\beta J)$: the divergence of the spin glass susceptibility
 occurs exactly at the  value $T_c$ given by (\ref{Tc}) where the
 distribution of local  fields becomes non-trivial. The typical
 correlation is $\exp\left(r 
 \int dJ P_{\al,\eps}(J) \log (\tanh^2 \beta J)\right)$, it 
 behaves as $\eps^{2 r}\ll \eps^{\alpha r}$ in the small $\eps$
 limit. This means that the average correlation $C_2(r)$ is 
 totally dominated by rare realizations:  its numerical estimate would 
 require an average over $O(1/\eps^{(2-\alpha)r})$ samples. 

In order to get around this problem, one must solve analytically the
one-dimensional L\'evy spin glass problem described in
(\ref{eq:levy1d}). This can be done either with the replica approach
of \cite{MonWei}, or using a cavity type approach. Both methods give
the same result, the detailed computations will be given in
\cite{JEM}. Let us just describe in a nutshell the basic steps of the
cavity approach. 
One first  solves the one dimensional spin glass
model (\ref{eq:levy1d}) using  the cavity method. The solution is
given in terms of some cavity fields $g_n$ which satisfy the update
equations 
$g_{n+1}=h_{n+1}+u(g_n,J_n)$. Then one studies the spin glass correlation
through the  response of $g_n$ to a perturbation in $g_1$.  Calling
$\Delta_n=(\partial g_n/\partial g_1)^2$, linear response theory
gives $\Delta_{n+1}= (\partial u(g_n,J_n)/\partial g_n)^2
\Delta_n$. Let us denote by 
$P_n(g_n,\Delta_n)$ the joint probability distribution of $g_n$ and
$\Delta_n$, over realizations of the random variables $\{h_p\},\;
p\in\{1,n\}$,  and $\{J_p\},\; p\in\{1,n-1\}$. The update equations
giving $g_{n+1}$ and $\Delta_{n+1}$ in terms of $g_n$ and $\Delta_n$
induces  a mapping $P_{n+1}=F(P_n)$ for the joint probability 
distribution. In order to study this mapping, one can introduce the
function $f_n(g_n)= \int d\Delta_n \Delta_n P_n(g_n,\Delta_n)$. 
It satisfies the recursion relation:
\begin{equation}
f_{n+1}(g_{n+1})=\!\!\int \!\! d g_n \!\!\int\!\! dJ_n P_{\al,\eps}(J_n) 
 \!\!\int\!\! dh_{n+1}
P(h_{n+1})  \left(\frac{\partial u(g_n,J_n)}{dg_n}\right)^2 \!\!
f_n(g_n)\,\delta\left( g_{n+1}-[h_{n+1}+u(g_n,J_n)]\right)\ .
\end{equation}
This  linear  equation, $f_{n+1}(g_{n+1})=\int  dg_n
K(g_{n+1},g_n)f_n(g_n)$, defines the transfer matrix operator
$K(x,y)$. The correlation length $\xi$ is given in terms of the
largest eigenvalue $\nu$  of $K$ by $\nu=e^{-1/\xi}$. The
computation of $\nu$ is most easily done by changing from the right to
the left eigenvalue equation. This gives the eigenvalue equation: 
\begin{equation}
\nu \phi(x)=\int dJ P_{\al,\eps}(J) \int dh P(h)\left(\frac{\partial
 u(x,J)}{\partial x}\right)^2  \phi(h+u(x,J))\big)= \int dy\;
 K^T(x,y)\, \phi(y) 
\label{eq:phiupdate}
\end{equation}

The largest eigenvalue of the linear operator $K$ can be found
numerically  by iterating (\ref{eq:phiupdate}) $\phi_{n}(x)=\int dy
K^T(x,y)\phi_{n-1}(y)/Z_n$, starting from an
arbitrary  function $\phi_0(x)$. At each step the constant $Z_{n}$ is 
computed by imposing a normalisation  condition $\int
dx\phi_n(x)=1$. After many iterations the function $\phi_n(x)$ 
converges to the eigenvector of $K$ with the largest eigenvalue, and
the normalisation  converges to $\lim_{n\to \infty} Z_n=
\nu=\exp(-1/\xi)$. 

\begin{figure}
  \begin{minipage}[c]{.5\textwidth}
  \includegraphics[width=\textwidth]{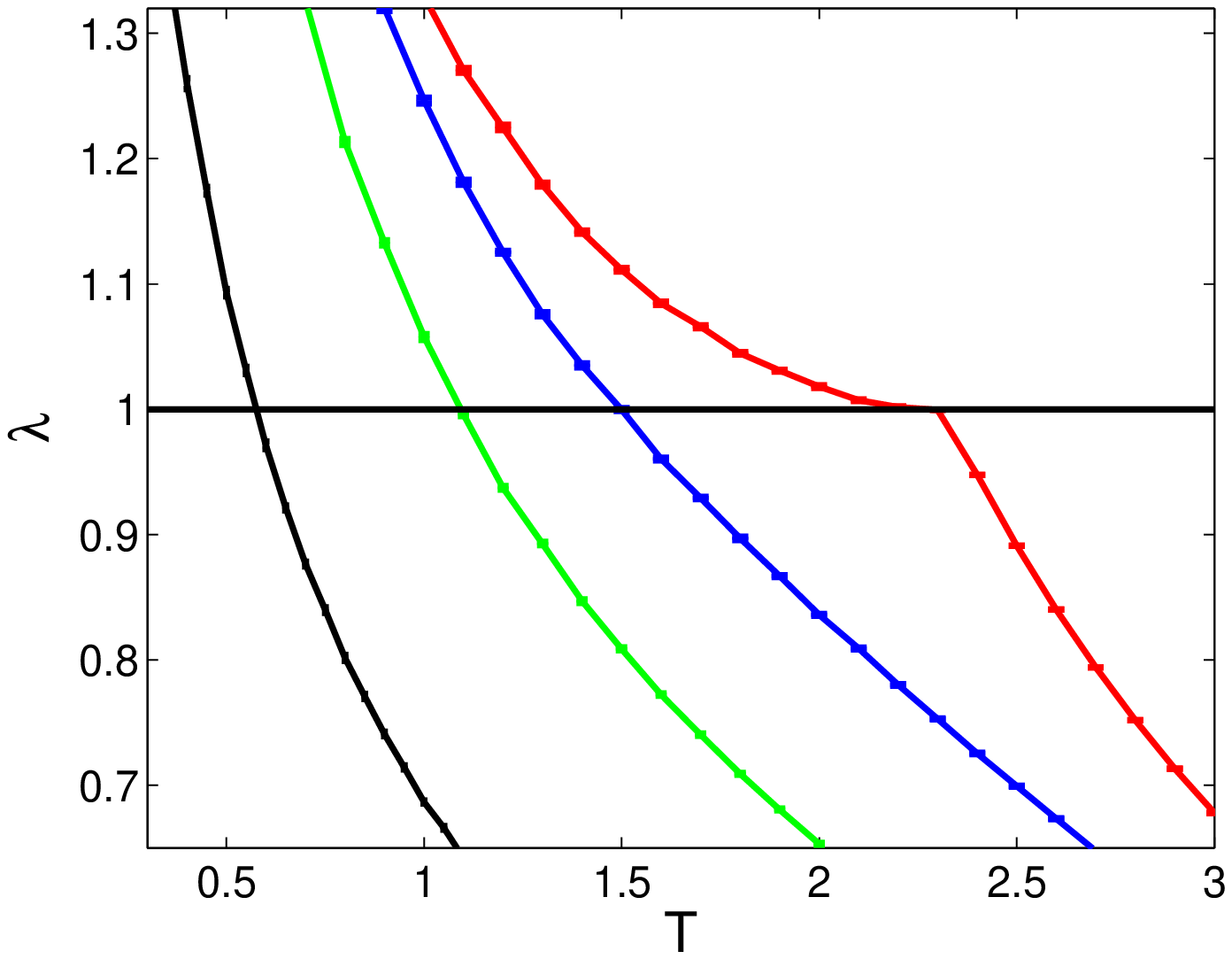}
\end{minipage}%
 \begin{minipage}[c]{.5 \textwidth}
 \includegraphics[width=\textwidth]{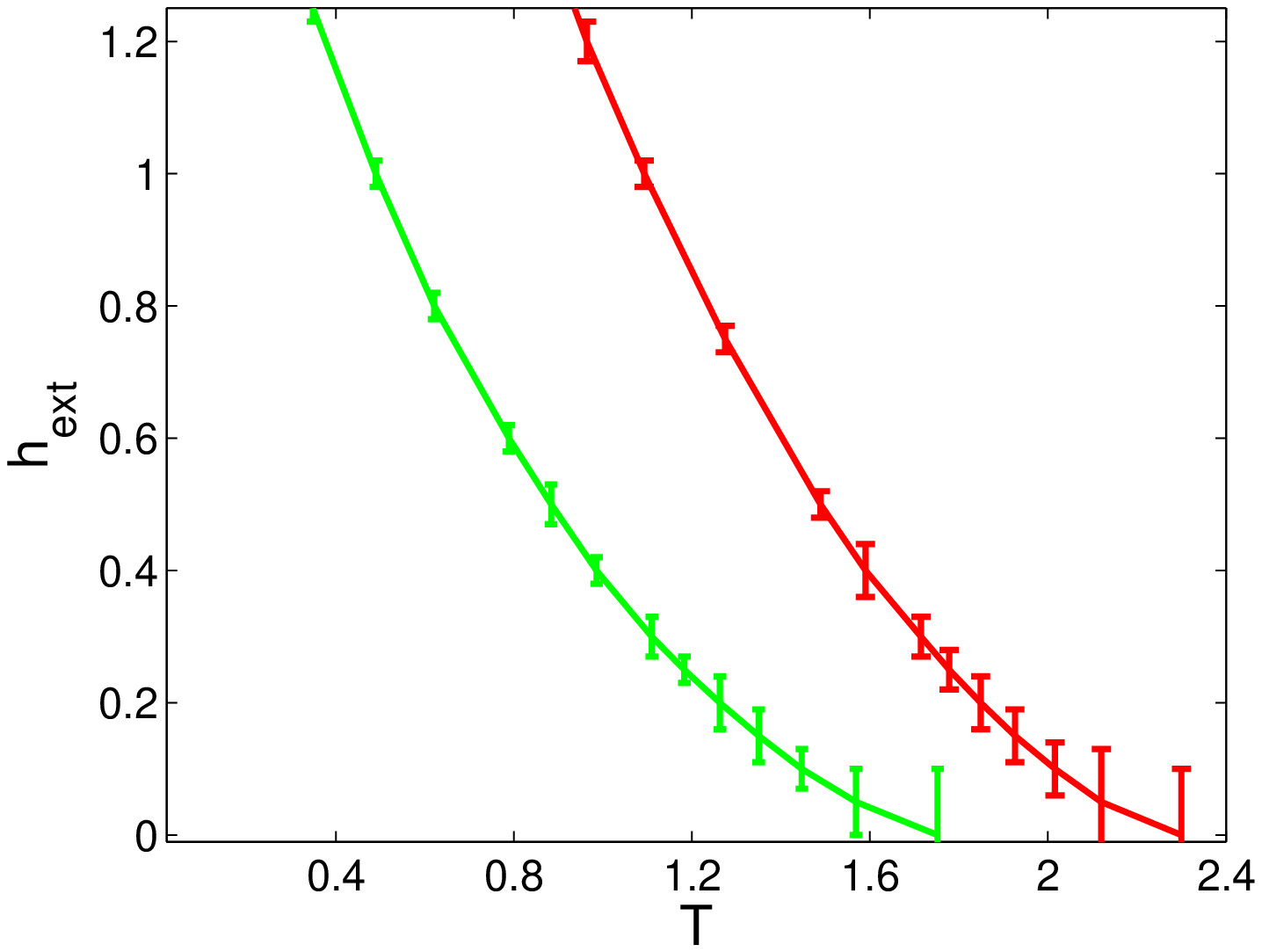}
\end{minipage}
  \caption{Determination of the de Almeida Thouless line. Left:
  Stability parameter $\lambda$ as functions of temperature for
  $\al=1.5$ and $\hex=0,\, 0.5,\, 1$ and $2$ (from right to
  left). From the intersection of the curves with the stability
  boundary $\lambda=1$ the AT-line is determined. Right: Phase diagram
  of a L\'evy spin glass with $\alpha=1.5$ (red) and $\alpha=1.1$
  (green). Above the AT-lines shown RS is stable, below it is
  unstable.}  
\label{fig:AT}
\end{figure}

In order to find the AT line one must hence use the $P(h)$
distribution as determined above with (\ref{hlargesmall}) and then
find the correlation length $\xi$ of the one-dimensional problem using
the $\phi_n$ iteration in (\ref{eq:phiupdate}). With this procedure
the limit $\eps \to 0$ is smooth, and this allows for  
a clean determination of the AT line, as shown in Fig.\ref{fig:AT}.  

The behaviour of the stability parameter $\lambda$ can be studied
analytically in zero external field close to the critical
temperature. Writing $\tau=1-T/T_c(\alpha)$, one must compute the
second and fourth moments of 
$P(h)$ up to order $\tau^2$, and then expand the eigenvalue equation 
(\ref{eq:phiupdate}). One finds after some work $\lambda=1+  (\alpha^2/3)
\; (T_2(\alpha)+2 T_4(\alpha))/ (T_2(\alpha)-T_4(\alpha))\; \tau^2 +
O(\tau^3)$, where $T_n(\alpha)= \int_0^{\infty} \al dx/x^{1+\alpha}
\tanh^{n} x$. As the coefficient of $\tau^2$ is 
positive for all $\alpha\in ]1,2[$,  the RS solution is always
unstable close to $T_c$, contrary to what was found with the Gaussian
Ansatz \cite{CB}. The same is obtained numerically in presence of an
external field: we have not found any evidence for a stable RS spin
glass phase, at all the values of $\alpha$ and $\hex$ that we have
studied.  

To summarize, we have shown how the L\'evy spin glass problem can be
studied naturally within the framework of diluted spin glasses, using
a decomposition of the couplings into strong and weak. The resulting
phase diagram is very similar to the one found in other mean field
spin glasses. In particular, the spin glass phase is never replica
symmetric. The large fluctuations due to the presence of rare
strong couplings request the introduction of some rather sophisticated
methods in order to compute the spin glass instability. These
fluctuations 
are even more pronounced in the case $\al<1$ not treated here where
the free energy ceases to be self-averaging. They will also complicate
the discussion of the RSB solution of the L\'evy glass.

Acknowledgements: We would like to thank Martin Weigt for interesting
discussions. Financial support from the Deutsche Forschungsgemeinschaft
under EN 278/7 is gratefully acknowledged. MM thanks the Alexander von 
Humboldt foundation. While we were writing up this work, the preprint
\cite{neri} has appeared where similar issues were addressed.

% generated by bibtex
%\bibliographystyle{apsrev}  
%\bibliography{levy}

\end{document}